%% file: paper.tex
\documentclass[sigconf]{acmart}
\usepackage{epsfig,endnotes}
\usepackage{multirow}
\usepackage{graphicx}
\usepackage{subfig}
\usepackage{grffile}

\newcounter{subcopyrightbox@save}
\usepackage{caption}
\usepackage{color, url}
\usepackage{xspace} 
\usepackage{mathrsfs}
\usepackage{amsmath}
\usepackage{epstopdf}
\usepackage{balance}
\usepackage{bm}
\usepackage{makecell}
\usepackage{cleveref}

\usepackage{algorithm}
\usepackage{algorithmic}

\copyrightyear{2021} 
\acmYear{2021} 
\acmConference[AsiaCCS '21]{2021 ACM ASIA Conference on Computer and Communications Security}{June 7--11, 2021}{Hong Kong, China}
\acmBooktitle{2021 ACM ASIA Conference on Computer and Communications Security (AsiaCCS '21), June 7--11, 2021, Hong Kong, China}
\acmPrice{15.00}
\acmDOI{10.1145/xxxxxxx.xxxxxxx}
\acmISBN{978-x-xxxx-xxxx-x/YY/MM}
\newcommand{\myparatight}[1]{\smallskip\noindent{\bf {#1}:}~}

\graphicspath{ {figs/} }

\begin{document}
\date{}

\title{Robust and Verifiable Information Embedding Attacks to Deep Neural Networks via Error-Correcting Codes}

\author{Jinyuan Jia, Binghui Wang, Neil Zhenqiang Gong}
\affiliation{%
  \institution{Duke University}
}
\email{{jinyuan.jia, binghui.wang, neil.gong}@duke.edu}


\input{abstract}



\maketitle

\input{intro}

\input{problem}

\input{encode}

\input{exp}

\input{casestudy}

\input{related}

\input{conclusion}

\begin{acks}
We thank the anonymous reviewers for insightful reviews. 
This work was supported by NSF grant No. 1937786. 
\end{acks}

{
\balance{
\bibliographystyle{plain}
\bibliography{refs,refs1,refs2}
}}

\end{document}

%% file: abstract.tex
\begin{abstract}
In the era of deep learning, a user often leverages a third-party machine learning tool to train a deep neural network (DNN) classifier and then deploys the classifier as an end-user software product (e.g., a mobile app) or a cloud service. In an information embedding attack, an attacker is the provider of a malicious third-party machine learning tool. The attacker embeds a message into the DNN classifier during training and recovers the message via querying the API of the black-box classifier after the user deploys it. Information embedding attacks have attracted growing attention because of various applications such as watermarking DNN classifiers and compromising user privacy. State-of-the-art information embedding attacks have two key limitations: 1) they cannot verify the correctness of the recovered message, and 2) they are not robust against post-processing (e.g., compression) of the classifier.

In this work, we aim to design information embedding attacks that are verifiable and robust against popular post-processing methods. Specifically, we leverage Cyclic Redundancy Check to verify the correctness of the recovered message. Moreover, to be robust against post-processing, we leverage Turbo codes, a type of error-correcting codes, to encode the message before embedding it to the DNN classifier. In order to save queries to the deployed classifier, we propose to recover the message via adaptively querying the classifier. Our adaptive recovery strategy leverages the property of Turbo codes that supports error correcting with a partial code. We evaluate our information embedding attacks using simulated messages and apply them to three applications (i.e., training data inference, property inference, DNN architecture inference), where messages have semantic interpretations. We consider 8 popular methods to post-process the classifier. Our results show that our attacks can accurately and verifiably recover the messages in all considered scenarios, while state-of-the-art attacks cannot accurately recover the messages in many scenarios.

\end{abstract}

%% file: intro.tex
\section{Introduction}

Deep neural networks (DNNs) have transformed multiple fields such as Computer Vision, Natural Language Processing, and Speech Processing. 
Suppose a user has a private training dataset and aims to learn a DNN classifier. The user often  leverages a third-party machine learning tool--such as Keras~\cite{Keras}, TensorFlow~\cite{abadi2016tensorflow}, and PyTorch~\cite{pytorch}--to train the DNN classifier. 
After training the classifier, the user could deploy it as an end-user software product, e.g., a mobile app or an Alexa skill; and the user could also deploy the classifier as a cloud service, e.g., the user can upload the classifier to Google AI Platform~\cite{google} or IBM Watson Machine Learning~\cite{IBM} for other customers to query the classifier.  
  
In an \emph{information embedding attack}, an attacker aims to \emph{embed} a \emph{message} in the classifier during training and \emph{recover} the message via querying the API of the black-box classifier after the user deploys it. The attacker could be the provider of the machine learning tool that the user uses to train its classifier; or the attacker could revise an existing machine learning tool and republishes it, which is downloaded and used by the user. Information embedding attacks can be used to watermark DNN classifiers~\cite{uchida2017embedding,rouhani2018deepsigns,adi2018turning,zhang2018protecting} and compromise user privacy when representing the user's private information (e.g., private training examples, sensitive property of the training dataset, or DNN architecture) as the message.

Without loss of generality, we assume the message is a bitstream.
State-of-the-art information embedding attacks~\cite{song2017machine} encode the bitstream message using a list of labels. Specifically, if the DNN classifier has $c$ classes, then each label can encode $\lfloor \log c \rfloor$ bits. Therefore, the attacker splits the bitstream message into $\lfloor \log c \rfloor$-bit blocks and encodes each block using the corresponding label. Then, the attacker generates some random data points using a pseudo-random number generator with a certain seed and assigns the labels to them. For convenience, we call these data points and their assigned labels \emph{attack dataset}.  To embed the message into the DNN classifier, the attacker augments the user's training dataset with the attack dataset and then uses standard techniques (e.g., Stochastic Gradient Descent) to train the classifier. After the user deploys the classifier, the attacker re-generates the attack data points using the pseudo-random number generator with the same seed, queries the classifier to get their labels, and recovers the bitstream message.

The state-of-the-art information embedding attacks~\cite{song2017machine} have two key limitations. First, they cannot verify whether the recovered message is correct or not. Second, they are not robust against post-processing of the DNN classifier. For instance, when the user aims to deploy the DNN classifier on resource-constrained devices such as mobile phone and IoT device, the user may post-process (e.g., prune certain weights) the classifier to reduce its size. As our experimental results show, state-of-the-art information embedding attacks are highly unlikely to accurately recover the message when the classifier is post-processed.

We aim to address these limitations in this work. Specifically, we leverage Cyclic Redundancy Check (CRC) to provide verifiability and error-correcting codes to be robust against post-processing. Our information embedding attacks embed the bitstream message into the DNN classifier through multiple steps. In Step I, we compute a CRC checksum of the message and append it to the message. In Step II, we encode the bitstream with checksum using a Recursive Systematic Convolutional (RSC) code based Turbo code (a type of error-correcting codes). In Step III, we generate an attack dataset to represent the encoded bitstream. In Step IV, we use the attack dataset to augment the training dataset to train the DNN classifier. 

A naive method to recover the message is to query the labels of all data points in the attack dataset (like state-of-the-art information embedding attacks), construct the encoded bitstream, and decode the encoded bitstream to obtain the message using the decoder of the Turbo code.  However, such method incurs a large amount of queries. This is because the attack dataset has a large number of data points due to encoding the message using error-correcting codes. When the user deploys the classifier as a cloud service, querying the cloud service often incurs economic cost, e.g., IBM Watson Machine Learning charges \$$5\times 10^{-4}$ per query.  Therefore, more queries incur larger cost to the attacker. To address the challenge, we propose to recover the message via \emph{adaptively} querying the classifier. Our method leverages the property of RSC-based Turbo codes, which supports error-correcting with a subsequence of the encoded bitstream. Specifically, we iteratively query the labels of the data points in the attack dataset; we recover the message using the current subsequence of the encoded bitstream; we check the correctness of the recovered message using CRC; and we query more labels only if the recovered message fails integrity checking.

We first evaluate our information embedding attacks using simulated messages on the CIFAR10 dataset with a ResNet neural network. We consider 8 popular post-processing methods such as fine tuning~\cite{adi2018turning}, pruning weights~\cite{han2015learning}, and pruning filters~\cite{li2016pruning}.  Then, we show three applications of information embedding attacks, i.e., \emph{training data inference} (using the FaceScrub dataset)~\cite{membershipInfer}, \emph{property inference} (using the US Census Income dataset)~\cite{Ateniese15,Ganju18,Melis19}, and \emph{neural network architecture inference} (using the CIFAR10 dataset), in which the messages have semantic interpretations. To perform training data inference (or property inference or neural network architecture inference), the attacker represents the training examples (or property of the training dataset or neural network architecture) as the message and embed it into the DNN classifier during training. After the user deploys the classifier, the attacker recovers the message via querying the classifier. Our results show that, compared to state-of-the-art information embedding attacks,  our attacks can accurately and verifiably recover the messages for all 8 post-processing methods with at most 1\% more accuracy loss of the DNN classifier (except one case) and 1-5 times more queries, which are acceptable. However, state-of-the-art attacks fail to correctly recover the messages in many cases.

Our contributions can be summarized as follows: 

\begin{itemize}
\item We propose error-correcting codes based information embedding attacks that are verifiable and robust against popular post-processing methods.

\item We propose to adaptively recover the message via leveraging the property of RSC-based Turbo codes that supports error correcting with a partial code.

\item We evaluate our information embedding attacks using simulated messages and three applications. 

\end{itemize}

%% file: problem.tex
\section{Problem Formulation}

\subsection{Threat Model}
We consider the same threat model as previous work~\cite{song2017machine}. 
In particular, we have two parties, i.e., \emph{user} and \emph{attacker}.

\myparatight{User} A user aims to train a deep neural network classifier and deploys it as an end-user AI software product or as a cloud service. 

{\bf Training classifier.} Suppose the user has a private training dataset. However, the user has limited machine learning expertise, and thus it is challenging for the user to develop its own tool to learn the neural network classifier. We consider the user leverages a third-party machine learning tool--such as Keras~\cite{Keras}, TensorFlow~\cite{abadi2016tensorflow}, and PyTorch~\cite{pytorch}--to train the classifier {locally}. For example, Keras~\cite{Keras} provides high-level APIs to learn neural networks; TensorFlow~\cite{abadi2016tensorflow}  provides many popular machine learning models which can be used without much efforts.

{\bf Deploying classifier.} After training the classifier, the user verifies the accuracy of the classifier using a validation dataset. If the classifier's performance is unsatisfying, then the user could discard the classifier. Otherwise, the user  
could deploy the classifier as an end-user software product (e.g., a mobile app, an Alexa skill) or as a cloud service (e.g., on Google AI Platform or IBM Watson Machine Learning) for other customers to query. When deploying the classifier as an end-user software on resource-constrained devices such as mobile phone and IoT device, the user may need to reduce the size of the classifier via \emph{model compression} techniques such as \emph{pruning}~\cite{han2015learning,li2016pruning,liu2018fine}. When deploying the classifier as a cloud service, customers may be charged based on the number of queries. For instance, IBM Watson Machine Learning charges \$$5\times 10^{-4}$ per query.

\myparatight{Attacker} An adversary aims to \emph{embed} some message into the classifier during training and \emph{recover} the message via querying the classifier after the classifier is deployed. We call this attack \emph{Information Embedding Attack}. 

{\bf Embedding message.} We assume the attacker can manipulate the training process to embed the message. Specifically, the attacker could be the provider of the machine learning tool that the user uses to train its classifier; or the attacker could revise an existing machine learning tool and republishes it, which is downloaded and used by the user. Since the user verifies the classifier's accuracy before deploying it, the attacker's goal is to embed the message during training such that the classifier's accuracy is not significantly affected for normal testing examples. 

We assume the machine learning tool cannot directly send the message to the attacker during training. For example, the user can train the classifier on machines that are disconnected with the Internet or monitor Internet usage such that the machine learning tool cannot send the message to the remote attacker via Internet.

{\bf Recovering message.} After the user deploys the classifier (the user may post-process the classifier using model compression techniques before deploying it) as an end-user software or a cloud service, the attacker recovers the message via querying the API of the classifier. In particular, we consider the attacker only uses the labels predicted by the API to recover the message, which is generally applicable without requiring access to the classifier's model parameters or the predicted confidence scores of queries.

\myparatight{Applications of information embedding attacks}
An attacker can use information embedding attacks to compromise confidentiality/privacy of the user's training data or classifier. For instance, the following shows three basic application scenarios of information embedding attacks. 

{\bf Training data inference.} The attacker can use information embedding attacks to infer the training data. 
Specifically, the attacker can encode some training examples that the attacker is interested in as the message. The attacker embeds the training examples into the neural network classifier and recovers them via querying the classifier. The inferred training examples could be further used to break face and voice based authentication systems~\cite{sharif2016accessorize} by using the training examples to synthesize faces and voices. Note that such training data inference is different from model inversion~\cite{fredrikson2015model} and  membership inference~\cite{membershipInfer,Nasr18}. Specifically, given access to a classifier's prediction API, model inversion is able to infer the aggregate training data in a particular class, e.g., to infer an aggregated recognizable face image for a particular person in a face recognition system; while membership inference aims to test whether a \emph{given} example is in the classifier's training data or not.

{\bf Property inference.} Rather than inferring specific training examples, the attacker may be interested in certain properties of the training dataset, which are known as \emph{property inference attacks}~\cite{Ganju18,Ateniese15}. Such property could include average age (e.g., 25), gender bias (e.g., 65\% Male vs. 35\% Female), and race distribution (e.g., 80\% White) of the individuals in the training dataset. To perform such property inference attacks, the attacker can encode a certain property of the training dataset as the message in an information embedding attack.  Existing property inference attacks~\cite{Ganju18,Ateniese15} require access to the model parameters and/or are effective for simple models (e.g., SVM, fully connected neural networks), while information embedding attacks based property inference only requires black-box access to the model and is applicable to complex deep neural networks.  

{\bf DNN architecture inference.} The attacker can also apply information embedding attacks to steal the architecture of the DNN classifier. Specifically, the attacker encodes the neural network architecture as the message in an information embedding attack.  The neural network architecture may be obtained by \emph{neural architecture search} techniques~\cite{zoph2016neural} for the given user's training dataset. After stealing the architecture, an attacker can further infer the model parameters via querying the model through model parameter stealing attacks~\cite{tramer2016stealing}, which subsequently enables the attacker to construct adversarial examples to evade the model. We note that neural network architecture is a hyperparameter of the neural network model.  Wang and Gong~\cite{WangHyper18} proposed attacks to steal hyperparameters in machine learning. However, their methods can only steal the regularization hyperparameters that are used to balance between a loss function and the regularization terms.

\subsection{Problem Definition}

We denote the message as ${m}$. Without loss of generality, we assume the message is a \emph{bitstream}. 
 We characterize an information embedding attack using two functions, i.e., \emph{Embed} and \emph{Recover}. The Embed function is executed by an untrusted machine learning tool to embed the message into the neural network classifier during training, while the Recover function recovers the message via querying the (post-processed) classifier. Formally, we define an information embedding attack as follows:

\begin{definition}[Information Embedding Attack]
Given a training dataset $D$ and a message $\mathbf{m}$. An information embedding attack consists of the following two functions: 
\begin{align}
    C &=Embed(D, \mathbf{m}) \\
    \mathbf{m}'&=Recover(API(C'))
\end{align}
where the Embed function outputs a machine learning classifier $C$ embedded with the message $\mathbf{m}$ and the  Recover function outputs the recovered message with an access to an API of the deployed classifier $C'$. The deployed classifier $C'$ may be a post-processed version of the classifier $C$.   
\end{definition}

An information embedding attack is to design the Embed and Recover functions.

\subsection{Design Goals}
We aim to design information embedding attacks that achieve the following goals.

\myparatight{Robust against post-processing} A user may post-process the model (e.g., pruning, fine-tuning) before deploying it on resource-constrained devices such as mobile phone and IoT device. Therefore, our information embedding attacks should be robust against post-processing techniques, i.e., our attacks should accurately recover the message even if the model is post-processed. 

\myparatight{Verifiable} Suppose the Recover function outputs a message ${m}'$. Due to the imperfection of the Embed function and noise/errors introduced by post-processing that the user applies to the classifier, the recovered message   ${m}'$ may not equal the original message ${m}$. Therefore, verifiability/integrity checking aims to enable the attacker to verify whether the message is exactly recovered or not.

\myparatight{Small impact on classifier accuracy} We measure the performance of the learnt classifier using its accuracy on testing examples. Embedding the message into the classifier should have a small impact on the classifier's accuracy.

\myparatight{Small number of queries} When the classifier is deployed as an end-user software, the attacker can issue an arbitrary number of queries to the classifier once the attacker obtains the software. When the classifier is deployed as a cloud service, the Recover function incurs costs due to querying the cloud service. Therefore, our attacks should require a small number of queries when the classifier is deployed as a cloud service.

The impact on the classifier accuracy and the number of queries represent the overheads of performing our information embedding attacks. 
Our goal is to design information embedding attacks that  are robust against post-processing, are verifiable, have small impacts on the classifier's accuracy, and require a small number of queries. Existing information embedding attacks~\cite{song2017machine} are not robust against post-processing and are not verifiable.

%% file: encode.tex
\section{Our Attacks}

\subsection{Overview}

Our key idea is to view an information embedding attack as a communications problem. Specifically, we view the Embed function that embeds the message to the classifier as the \emph{sender} of the message, and we view the Recover function as the \emph{receiver} of the message. The communications channel between the sender and the receiver is unreliable (i.e., the message may be altered during transmission) due to the imperfection of the Embed function and the noise/errors introduced by the post-processing that the user applies to the classifier before deploying it.  This perspective enables us to leverage coding techniques to design robust information embedding attacks.

\myparatight{Embed} Our Embed function consists of multiple steps. First, to enable integrity checking, we compute a 32-bit \emph{Cyclic Redundancy Check (CRC-32)}~\cite{peterson1961cyclic} checksum for the message bitstream and append the checksum to the bitstream. CRC is an error-detection code widely used in network communications and data storage systems. We note that cryptographic hash functions could also be used to compute a checksum and enable integrity checking. However, cryptographic hash functions have longer checksums (e.g., MD5 has 128-bit checksum), which makes it harder to embed the bitstream with checksum in the classifier.  Second, we encode the bitstream with checksum to another \emph{encoded bitstream} using \emph{Turbo code}~\cite{berrou1993near}. Turbo codes are popular error-correcting codes that can recover a bitstream even if certain amounts of bits are altered.  Turbo codes are used in mobile communications and satellite communications to correct the transmitted data, which may be altered due to the unreliable wireless communications. Third, we represent the encoded bitstream using labels of some random data points (called \emph{attack dataset}). Fourth, we use the attack dataset to augment the training dataset that is used to learn the model.

\myparatight{Recover} To save the number of queries, our Recover function recovers the message via \emph{adaptively} 
querying the API of the deployed classifier. Specifically, we first query the labels of some data points in the attack dataset,  transform the labels to a bitstream, and decode it using the decoder of the Turbo code. The decoded bitstream consists of the message bitstream and the CRC checksum. We use the CRC checksum to perform integrity checking. 
If the message is not recovered exactly, we query the labels of some other data points in the attack dataset. We repeat this process until recovering the exact message or querying the labels of all data points in the attack dataset.  

Next, we will first briefly review some background knowledge on coding that is used by our attacks. Then, we will describe our Embed and Recover functions.

\subsection{Background on Coding}

\myparatight{Cyclic Redundancy Check (CRC)~\cite{peterson1961cyclic}} CRC is an error-detection code widely used in communications and data storage to detect errors. Specifically, CRC computes a fixed-length checksum for a bitstream.  The checksum is appended to the bitstream. When verifying the integrity, we compute the CRC checksum of the bitstream again; and the bitstream maintains integrity with an overwhelmingly high probability if the checksum equals the one appended to the bitstream. In particular, if the bitstream is randomly corrupted, then CRC can detect the corruption with a probability $1-2^{-L}$, where $L$ is the length of the checksum.  In our experiments, we adopt CRC-32 whose checksum is always 32 bits.

\myparatight{Recursive Systematic Convolutional (RSC) codes~\cite{elias1955coding,berrou1993near}} RSC is a type of error-correcting codes. Suppose we have a bitstream $m$. Roughly speaking, RSC encodes the bitstream to another bitstream $m+ e_1+ e_2\cdots e_k$, where each $e_i$ is a bitstream that has the same length with the bitstream $m$, the symbol $+$ means concatenation of bitstreams, and $k+1$ is the encoding ratio that is a tunable parameter of RSC. Note that the original bitstream is the first sub-block of the encoded bitstream. A larger encoding ratio can tolerate a larger fraction of errors but leads to a longer encoded bitstream. 

Suppose the encoded bitstream is corrupted to be $m'+ e_1'+ e_2'\cdots e_k'$. RSC has a decoder $D$ which recovers the original bitstream $m$ from the corrupted encoded bitstream, i.e., $m=D(m'+ e_1'+ e_2'\cdots e_k')$, when the errors in the encoded bitstream are bounded. We note that any sub-block $m'+ e_1'+ e_2'\cdots e_i'$ of the encoded bitstream can also be used to recover the original bitstream $m$, where $i=1, 2,\cdots k$. Specifically, such a sub-block corresponds to an RSC code with an encoding ratio of $i+1$ and a decoder $D_{i+1}$ that decodes the bitstream as   $m=D_{i+1}(m'+ e_1'+ e_2'\cdots e_i')$. We leverage this observation to adaptively recover the message.

\myparatight{Turbo codes~\cite{berrou1993near}} Turbo codes are another type of error-correcting codes. Turbo codes can tolerate more errors than RSC, given the same encoding ratio.  
One category of Turbo codes is designed based on RSC. Specifically, given a bitstream $m$, Turbo code first uses an RSC code to get an encoded bitstream $m+ e_1+ e_2\cdots e_k$. Then, the Turbo code permutes the original bitstream to be $\bar{m}$ and uses the same RSC code to encode the permuted bitstream, which generates another encoded bitstream $\bar{m}+ \bar{e}_1+ \bar{e}_2\cdots \bar{e}_k$. The Turbo code concatenates the two encoded bitstreams (except the permuted bitstream $\bar{m}$) as final encoded bitstream, i.e., the final encoded bitstream is   $m+ e_1\cdots e_k+\bar{e}_1\cdots \bar{e}_k$.  The permuted bitstream $\bar{m}$ is not included in the final encoded bitstream as it is simply redundant.  The encoding ratio of the Turbo code is $2k + 1$. 

Suppose the encoded bitstream is randomly corrupted to be $m'+ e_1'\cdots e_k'+\bar{e}_1'\cdots \bar{e}_k'$. The Turbo code has a decoder $T$ that can recover the original bitstream from the corrupted encoded bitstream, i.e., $m=T(m'+ e_1'\cdots e_k'+\bar{e}_1'\cdots \bar{e}_k')$, when the errors are bounded. We note that, like the RSC code, any sub-block $m'+ e_1'\cdots e_i'+\bar{e}_1'\cdots \bar{e}_i'$ of the encoded bitstream corresponds to a Turbo code with an encoding ratio of $2i+1$ and can be used to recover the original bitstream using the corresponding decoder $T_{2i+1}$. We will leverage this observation to adaptively recover the message in our Recover function to save queries to the classifier's API.

\subsection{Embed Function}
The Embed function is essentially the machine learning tool that trains a classifier for the user. Our Embed function consists of four steps, which we describe one by one.  Step I enables verifiability, Step II enables robustness, while Step III and Step IV embed the message into the neural network.

\myparatight{Step I} We compute a CRC-32 checksum for the message bitstream $m$ and append the checksum to the bitstream. We denote the bitstream with checksum as $m_c=m+cs$, where $cs$ is the checksum.  This step is to support integrity checking.

\myparatight{Step II} We encode the bitstream $m_c$ using a Turbo code with an encoding ratio $2k+1$. For convenience, we denote the encoded bitstream as $m_e=m_c + e_1\cdots e_k+\bar{e}_1\cdots \bar{e}_k$. 

\myparatight{Step III} We represent the encoded bitstream using labels of some random data points, which we call \emph{attack dataset}.  
Suppose the classifier has $c$ possible labels. Therefore, each label can represent  a $\lfloor \log c \rfloor$-bit bitstream. For instance, if the classifier has 10 possible labels, then each label can represent a 3-bit bitstream. Specifically, label 0 represents  $000$, label 1 represents $001$, and label 7 represents $111$. Note that label 8 and 9 are not used to represent bitstreams in the example.  We represent the encoded bitstream using an {attack dataset} with carefully designed labels. Specifically, we divide the encoded bitstream $m_e$ as $\lfloor \log c \rfloor$-bit sub blocks (we append 0 to the last sub-block if needed). We represent the $i$th sub block using a label $l_i$. Suppose we have $n$ such blocks. We generate $n$ random data points using a pseudo-random number generator with a certain seed. Furthermore, we assign label $l_i$ to the $i$th data point. The random data points and their assigned labels are our attack dataset.

\myparatight{Step IV}  We use the attack dataset to augment the user's training dataset when training a neural network classifier. The machine learning tool uses standard \emph{Stochastic Gradient Descent} to learn the classifier. Specifically,  in each iteration, we sample a small batch from both the original training dataset and our attack dataset, and combine them together to update the model parameters. The learnt classifier would predict $l_i$ as the label of the $i$th data point in the attack dataset, even if the data points are randomly generated. This is because neural networks have large capacity, which can represent more information other than the main classification task~\cite{zhang2016understanding}.

\myparatight{Comparison with existing information embedding attacks~\cite{song2017machine} and watermarking~\cite{uchida2017embedding,rouhani2018deepsigns,zhang2018protecting,adi2018turning}} Existing information embedding attacks~\cite{song2017machine} only have Step III and Step IV. Specifically, in Step III, they directly represent the message bitstream $m$ using an attack dataset with carefully crafted labels. As a result, they cannot support integrity checking and are not robust against post-processing of the learnt classifier. 
In watermarking, a model owner embeds a watermark to a neural network during training and verifies the watermark for another deployed model that is potentially a pirated version of the model owner's neural network. One category of watermarking techniques~\cite{zhang2018protecting,adi2018turning}, which only require access to the deployed model's prediction API to recover the watermark, treat some data points with certain labels (i.e., attack dataset in our terminology) as watermark and use them to augment the training dataset when training the neural network.  
These watermarking techniques~\cite{zhang2018protecting,adi2018turning} essentially only use Step IV.

\begin{algorithm}[t]
    \caption{{Recover} Function}
    \begin{algorithmic}[1]
    \REQUIRE Groups of data points in our attack dataset, and API of the classifier. \\
    \ENSURE  Message bitstream $m$. \\
         \STATE $m_c'=m'+cs' \longleftarrow$ Labels of the first group \;
         \IF {CRC-32($m'$) == $cs'$}
         \RETURN $m'$ \;
         \ENDIF
         \STATE i= 1 \;
         \WHILE {$i \leq k$}
         \STATE $e_i' \longleftarrow$ Labels of the $2i$th group \;
         \STATE //Apply RSC decoder \;
         \STATE $m'+cs' \longleftarrow D_{i+1} (m_c'+e_1'\cdots e_i') $ \;
         \IF {CRC-32($m'$) == $cs'$}
         \RETURN $m'$ \;
         \ENDIF
         \STATE $\bar{e}_i' \longleftarrow$ Labels of the $(2i+1)$th group \;
         \STATE //Apply Turbo decoder \;
         \STATE $m'+cs' \longleftarrow T_{2i+1} (m_c'+e_1'\cdots e_i'+\bar{e}_1'\cdots \bar{e}_i') $ \;
         \IF {CRC-32($m'$) == $cs'$}
         \RETURN $m'$ \;
         \ENDIF
         \STATE $i \longleftarrow i + 1$ \;
         \ENDWHILE \;
         \RETURN $m'$ \;
\end{algorithmic}
    \label{algorithm2}
\end{algorithm}

\subsection{Recover Function}

The {Recover} function aims to recover the message bitstream $m$ via querying the API of the deployed classifier. We consider both the Embed and Recover functions know the size of the message, the number of labels, and the seed of the pseudo-random number generator used to generate the data points in our attack dataset. Therefore, the Recover function can re-generate the data points in the attack dataset.  A naive implementation of the Recover function is to query the labels of all data points in the attack dataset, reconstruct the encoded bitstream $m_e$ (some bits may be corrupted due to imperfection of Embed and post-processing), and use the decoder of the Turbo code to decode the bitstream $m_c$, which can be used to perform integrity checking and recover the message. 
However, this naive implementation requires querying all data points in the attack dataset, which incurs large costs when the classifier is deployed as a cloud service.  

To save queries, we propose to adaptively recover the message bitstream and only make more queries if needed. Our key idea is to alternately apply decoders of the RSC and Turbo codes. Specifically, we divide the data points in the attack dataset into $2k+1$ groups: the labels of the first group represent the block $m_c$ of the encoded bitstream; the labels of the $(2i)$th group  represent the block $e_i$ of the encoded bitstream; and the labels of the $(2i+1)$th group represent the block $\bar{e}_i$, where $i=1,2,\cdots,k$. 

We first query the labels of data points in the first group, which transforms to a bitstream $m_c'$. We use CRC-32 to check the integrity of the bitstream $m_c'$ via treating the last 32 bits as the checksum. Specifically, we split $m_c'=m' + cs'$ and check whether $cs'=$CRC-32$(m')$. If yes, then we treat the bitstream $m'$ as the message. Otherwise, we query the labels of data points in the second group, which leads to a bitstream  $e_1'$ that may be a corrupted version of the bitstream $e_1$. We use an RSC decoder $D_2$ with an encoding ratio of 2 to decode the bitstream we obtained so far, i.e., $D_2(m_c' + e_1')$. Then, we check the integrity of the decoded bitstream using CRC-32. If the decoded bitstream passes the integrity checking, then we treat the decoded bitstream (except the last 32 bits) as the message. Otherwise, we continue querying the labels of data points in the third group, which leads to a bitstream $\bar{e}_1'$ that may be a corrupted version of the bitstream $\bar{e}_1$. Then, we use a Turbo decoder $T_3$ with an encoding ratio of 3 to decode the bitstream we have obtained so far, i.e., $T_3(m_c' + e_1' + \bar{e}_1')$. Then, we check the integrity of the decoded bitstream using CRC-32 and recover the message if it passes the integrity check. We repeat this process until the message is accurately recovered or we have queried the data points  in all groups.

Algorithm~\ref{algorithm2} shows the key components of our Recover function. Roughly speaking, after querying the labels of the data points in the $(2i)$th group, we use an RSC decoder with an encoding ratio of $i+1$ to decode a bitstream from the encoded bitstream $m_c'+e_1'\cdots e_i'$. If the decoded bitstream passes integrity check, then we recover the message from it. Otherwise, we continue querying labels of  the data points in the next group. After querying the labels of the data points in the $(2i+1)$th group, we use a Turbo decoder with an encoding ratio of $2i+1$ to decode a bitstream from the bitstream $m_c'+e_1'\cdots e_i'+\bar{e}_1'\cdots \bar{e}_i'$. If the decoded bitstream passes integrity check, then we recover the message. Otherwise, we continue querying labels of the data points in the next group.

%% file: exp.tex
\section{Evaluation}
\label{exp}

We first show evaluation results for some simulated messages. In the next section, we will show three applications of our information embedding attacks.

\subsection{Dataset and Neural Network Model}

We use the well known CIFAR10 dataset, which contains 60,000 color images (50,000 training images and 10,000 testing images, respectively) with $10$ classes. The size of each image is $32\times 32$. We consider the ResNet-18~\cite{He_2016_CVPR} neural network architecture for this dataset, which achieves 91\% classification accuracy in our experiments. {We note that our attacks are also applicable to other neural network architectures.}  We learn the model parameters using Stochastic Gradient Descent (SGD) to optimize the cross entropy loss. Specifically, we run 200 epochs and set batch size to 64. In information embedding attacks, for each batch of training dataset, we sample 10 data points from the attack dataset and add them to the batch. 
We set the  learning rate to 0.1 and  decrease it by a factor $0.1$ in the 150th, 170th, and 190th epochs for better convergence.

\subsection{Simulating a Message}
Without loss of generality, we consider the message includes 500 8-bit random integers, which translates to a random 4,000-bit bitstream. In the next section, we will show three applications of our attacks, where the messages have semantic interpretations.

\subsection{Compared Attacks}
We compare the following three information embedding attacks.

\myparatight{Direct Encoding (DE)~\cite{song2017machine}} This attack directly encodes the message using the labels of the data points in the attack dataset in the Embed function, and recovers the message via querying the labels of all data points in the attack dataset in the Recover function. 

\myparatight{Error-correcting Codes (ECC)} This is our attack. ECC appends a CRC-32 checksum to the message and encodes the message using a Turbo code with an encoding ratio of 5 in the Embed function. In the Recover function, ECC adaptively queries the labels and recovers the message.

\myparatight{Error-correcting Codes with all queries (ECC-All)} This is a variant of our attack. In particular, ECC-All uses the same Embed function with ECC. However, ECC-All queries the labels of all data points in the attack dataset in the Recover function. We compare ECC-All with ECC to demonstrate that ECC can save a large amount of queries in some scenarios.

\subsection{Evaluation Metrics}

\myparatight{Mean Squared Error (MSE) and Bit Error Rate (BER)} In our simulation, the message includes 500 integers. Therefore, we use MSE to measure the accuracy of the recovered integers. In general, the message can be represented as a bitstream. Therefore, we further consider BER as a metric. Specifically, BER is the fraction of incorrect bits in the recovered bitstream.   

\myparatight{Classification accuracy (ACC)} We measure the performance of a classifier using its classification accuracy on the testing dataset.  Our attacks have small impacts on the classifier's accuracy.

\myparatight{Number of queries} We also measure an attack using the required number of queries to the classifier. We stress that when the classifier is deployed as an end-user software product, the attacker can make arbitrary number of queries to the classifier. When the classifier is deployed as a cloud service, the attacker may be charged according to the number of queries. Therefore, a small number of queries corresponds to a small economic cost for the attacker when the classifier is deployed as a cloud service.

\begin{table*} [!t]\renewcommand{\arraystretch}{1.3}
    \centering
    \caption{Results of embedding/recovering a message that includes 500 random 8-bit integers (i.e., a random 4,000-bit bitstream). ECC-All and ECC have the same MSE, BER, and ACC, so we omit the corresponding results for ECC-All. Compared to the existing information embedding attack DE, our attack ECC can accurately and verifiably recover the message with at most $1\%$ more accuracy loss of the classifier and 1-5 times more queries.}
    \begin{tabular}{|c|c|c|c|c|c|c|c|c|c|c|c|c|} \hline
      Metric&
      Attack &
    No-Post & FTLL & FTAL&PRWT&PRWT-FT & PRFL & PRFL-FT & ADPR & ADPR-FT \\
      \hline
        \multirow{2}{*}{MSE}&DE&0&0&116&512&1,316&1,831&2,075&1,076&1,420  \\ \cline{2-11}
                            
                            &ECC&0&0&0&0&0&0&0&0&0  \\ \hline
        \multirow{2}{*}{BER (\%)}&DE&0&0&0.7&3.0&7.6&9.4&10.5&6.2&8.8   \\ \cline{2-11}
                           
                            &ECC&0&0&0&0&0&0&0&0&0  \\ \hline
              \multirow{2}{*}{ACC (\%)}&DE&90.10&89.76&90.41&87.10&89.99&87.21&89.52&87.64&89.68 \\ \cline{2-11}
                            
                            &ECC&89.88&89.75&90.22&87.01&89.44&87.04&89.14&87.61&89.49 \\ \hline
       \multirow{3}{*}{\#Queries}&DE&1,334&1,334&1,334&1,334&1,334&1,334&1,334&1,334&1,334  \\ \cline{2-11}
                                  &ECC&1,344&1,344&4,036&4,036&4,036&4,036&4,036&6,726&6,726  \\  \cline{2-11}
                                  &ECC-All&6,726&6,726&6,726&6,726&6,726&6,726&6,726&6,726&6,726 \\ \hline

    \end{tabular}
    \vspace{-2mm}
    \label{result-simulation}
\end{table*}

\subsection{Post-processing}
The user may post-process the learnt classifier  before  deploying it~\cite{han2015learning,han2015deep,li2016pruning}. We evaluate the following 8 post-processing methods that were considered in previous studies~\cite{adi2018turning,han2015learning,li2016pruning,liu2018fine}.

\myparatight{Two variants of fine tuning (FTLL and FTAL)~\cite{adi2018turning}} Following previous work~\cite{adi2018turning}, we consider two different ways to fine-tune a model. One way is to  fine-tune the parameters of the last neural network layer (denoted as FTLL), and the other is to fine-tune all layers (denoted as FTAL).
We also assume the user uses the entire training dataset to fine-tune the classifier.

\myparatight{Two variants of pruning weights (PRWT and PRWT-FT)~\cite{han2015learning}}\\When the user aims to deploy the classifier on resource-constrained devices such as mobile phone and IoT device, the user may use model compression methods to reduce the size of the classifier. Weight pruning is a popular method to compress models. In particular, PRWT prunes the weights of the neural network that have small magnitudes. 
 Pruning more weights makes the model smaller but also sacrifices the classification accuracy more. In our experiments, we prune the weights such that the classifier's classification accuracy is reduced by at most   $3\%$. Specifically, in PRWT, we first rank the weights according to their absolute values and iteratively remove the neural network connections with small weights. In each iteration, we prune $0.5\%$ of weights and re-calculate the classifier's accuracy on the testing dataset. We repeat the process until the classifier's accuracy is reduced more than $3\%$.   
In PRWT-FT, we further fine tune the remaining weights using the training dataset to improve the classifier's accuracy.

\myparatight{Two variants of pruning filters (PRFL and PRFL-FT)~\cite{li2016pruning}} Pruning weights with small magnitudes is very likely to prune connections in the fully connected layers. However, the convolutional layers require sparse BLAS libraries or specialized hardware for acceleration~\cite{iandola2016squeezenet}. As a result, pruning weights may not necessarily reduce the computation overhead as desired.  To address the challenge, researchers proposed to prune filters~\cite{li2016pruning} that have small impact on the neural network's output.
Specifically, in Prune Filters (PRFL), a filter's importance is measured by the $L_1$ norm of its weights~\cite{li2016pruning}.  ResNets have three stages of residual blocks for feature maps. For ResNet-18, each stage contains three residual blocks and each residual block contains two layers. Similar to~\cite{li2016pruning}, we consider pruning the first layer in each residual block as pruning the second layer would require pruning many other layers. 
For fully connected layers, we prune neurons and measure a neuron's importance using the $L_1$-norm of its weights.  

Like PRWT, we iteratively prune filters and neurons until the classifier's accuracy is reduced by more than $3\%$. Specifically, in each iteration, we go through each layer that is considered for pruning and prune $p\%$ of its filters/neurons; then we re-calculate the classifier's accuracy on the testing dataset. We set $p=\frac{1}{f}$, where $f$ is the smallest number of filters/neurons in a layer among the layers that are considered for pruning. We set $p$ in this way such that we prune at least one filter/neuron for a layer in each iteration.  PRFL-FT further fine tunes the remaining parameters obtained by PRFL using the training dataset.

\myparatight{Two variants of adapted pruning filters (ADPR and ADPR-FT)~\cite{liu2018fine}} 
Rather than using $L_1$-norm to measure the importance of a filter or neuron, Kang et al.~\cite{liu2018fine} proposed to measure the importance using the sum of activation values of a filter or neuron for the training dataset. Moreover, they proposed to prune the last convolutional layer as it could achieve a better tradeoff between model size and accuracy. This method was designed to remove backdoors in neural networks. We adapt this method to ResNet-18 (denoted as ADPR) to remove the messages. Specifically, we prune the first layer in the last residual block of the last stage in ResNet-18. We do not  prune the last convolutional layer because pruning the last convolutional layer requires pruning many other layers due to short cut connections for ResNet. Moreover, we prune neurons in the last fully connected layer. Like PRWT and PRFL, we iteratively prune filters and neurons until the classifier's accuracy is reduced by more than $3\%$. Specifically, in each iteration, we prune one filter in the considered convolutional layer and one neuron in the considered fully connected layer. 
ADPR-FT further fine tunes the pruned neural network using the training dataset.

When fine tuning is needed (e.g., FTLL, PRWT-FT, and ADPR-FT), we set epochs to $100$ and use a small learning rate, i.e., $0.001$.

\subsection{Results}

Table~\ref{result-simulation} shows the results for the compared attacks and different post-processing methods.

\begin{table}[!hbt]\renewcommand{\arraystretch}{0.9}
    \centering
    \caption{Architecture of the neural network for FaceScrub.}
    \begin{tabular}{|c|c|} \hline 
    Layer Type & Layer Parameters  \\ \hline
    \multicolumn{2}{|c|}{Input  $50\times 50$} \\ \hline
    Convolution& $32\times 3 \times 3$, strides=$(1, 1)$, padding=same  \\ 
    Activation& ReLU  \\ \hline
    Convolution& $32\times 3 \times 3$, strides=$(1, 1)$  \\ 
    Activation& ReLU  \\ 
    Pooling& MaxPooling$(4\times 4)$, padding=same  \\ \hline
    Convolution& $64\times 3 \times 3$, strides=$(1, 1)$, padding=same  \\ 
    Activation& ReLU  \\ \hline
    Convolution& $64\times 3 \times 3$, strides=$(1, 1)$  \\ 
    Activation& ReLU  \\ 
    Pooling& MaxPooling$(4\times 4)$, padding=same  \\ 
    Dropout& $0.2$  \\ \hline
    Fully Connected& 1024  \\ 
    Activation& ReLU  \\
    Dropout& $0.2$  \\ \hline
    Fully Connected& 530  \\ 
    Activation& softmax  \\ \hline
    \multicolumn{2}{|c|}{Output} \\ \hline
    \end{tabular} 
    \vspace{-2mm}
    \label{architecture} 
\end{table}

\myparatight{MSE and BER} Without post-processing, all compared attacks can accurately recover the message. We note that existing attack DE cannot verify whether the message is correctly recovered or not, while our attack can. However, when the classifier is post-processed, DE incurs large errors at recovering the message for 7 of the 8 post-processing methods, while our attack can still accurately recover the message. The reason is that our attack leverages error-correcting codes to correct the errors in the recovered message.

\myparatight{ACC} Compared to the existing attack DE, our attack incurs a larger loss of the classifier's accuracy. This is because our attack adopts error-correcting codes, which requires a larger attack dataset to encode the bitstream message. A larger attack dataset makes the learnt classifier less accurate on the normal testing dataset. However, compared to the existing attack DE, our attack reduces the accuracy by at most $1\%$ more for no post-processing and  all the 8 post-processing methods, which is acceptable.

\myparatight{Number of queries} Our attack ECC requires 1-5 times more queries than DE. Specifically, for no post-processing and FTLL, ECC just requires 10 more queries than DE. The 10 queries are used to recover the CRC-32 checksum that ECC uses to verify the correctness of the recovered message. ECC requires at most 5 times more queries than DE because we use a Turbo code with an encoding ratio of 5. ECC uses less queries than ECC-All in many scenarios because ECC adaptively recovers the message.   

In summary, compared to the existing attack DE, our attack ECC can \emph{accurately} and \emph{verifiably} recover the message with at most $1\%$ more accuracy loss of the classifier and 1-5 times more queries.

%% file: casestudy.tex
\section{Three Applications}
\label{applications}
We show three real-world applications of information embedding attacks, i.e., training data inference, property inference, and neural network architecture inference. We note that the experimental results in the previous section do not necessarily imply the success of these applications. Therefore, we evaluate our information embedding attacks in different applications to demonstrate the effectiveness of our attacks in practice. 


\begin{table*} [hbt]\renewcommand{\arraystretch}{1.2}
    \centering
    \caption{Results of training data inference. DE-Pixel concatenates the pixel values of the considered training image as the message; DE and ECC treat the binary JPEG image file as the message. MSE measures the error of the recovered pixel values, and BER is the bit error rate of the recovered bitstream message.}
    \begin{tabular}{|c|c|c|c|c|c|c|c|c|c|c|c|c|} \hline
      Metric&
      Attack &
       No-Post & FTLL & FTAL & PRFL &PRWT&PRWT-FT& PRFL-FT & ADPR & ADPR-FT \\
      \hline
        \multirow{3}{*}{MSE} &DE-Pixel &0 &0 &205&0&19&0&8&267&1,352   \\ \cline{2-11}
                            &DE &0 &0 &2&0&8&0&0&107&313  \\ \cline{2-11}
                            &ECC &0 &0 &0&0&0&0&0&0&0  \\ \hline
        \multirow{3}{*}{BER (\%)} &DE-Pixel &0 &0 &1.0&0&0.1&0&0.02&1.4&7.0   \\ \cline{2-11}
                            &DE &0 &0 &0.005&0&0.03&0&0&0.3&1.5  \\ \cline{2-11}
                            &ECC &0 &0 &0&0&0&0&0&0&0  \\ \hline
	       \multirow{3}{*}{ACC (\%)}&DE-Pixel &79.11&79.07&79.43&76.39&78.74&78.26&79.50&76.77&78.82  \\ \cline{2-11}
                            &DE&79.61&79.50&79.34&76.73&79.19&78.29&79.68&77.18&79.33 \\ \cline{2-11}
                            &ECC&79.35&79.19&79.79&76.39&79.19&78.04&79.75&76.87&79.23  \\ \hline
        \multirow{3}{*}{\#Queries}&DE-Pixel &12,000&12,000&12,000&12,000&12,000&12,000&12,000&12,000&12,000  \\ \cline{2-11}
                            &DE&4,135&4,135&4,135&4,135&4,135&4,135&4,135&4,135&4,135 \\ \cline{2-11}
                            &ECC&4,141&4,141&12,426&12,426&20,708&8,282&12,426&12,426&20,708  \\ \hline
    \end{tabular}
    \label{training_data_table}   
\end{table*}

\begin{figure*}[hbt]
    \centering
    {\includegraphics[width=\textwidth]{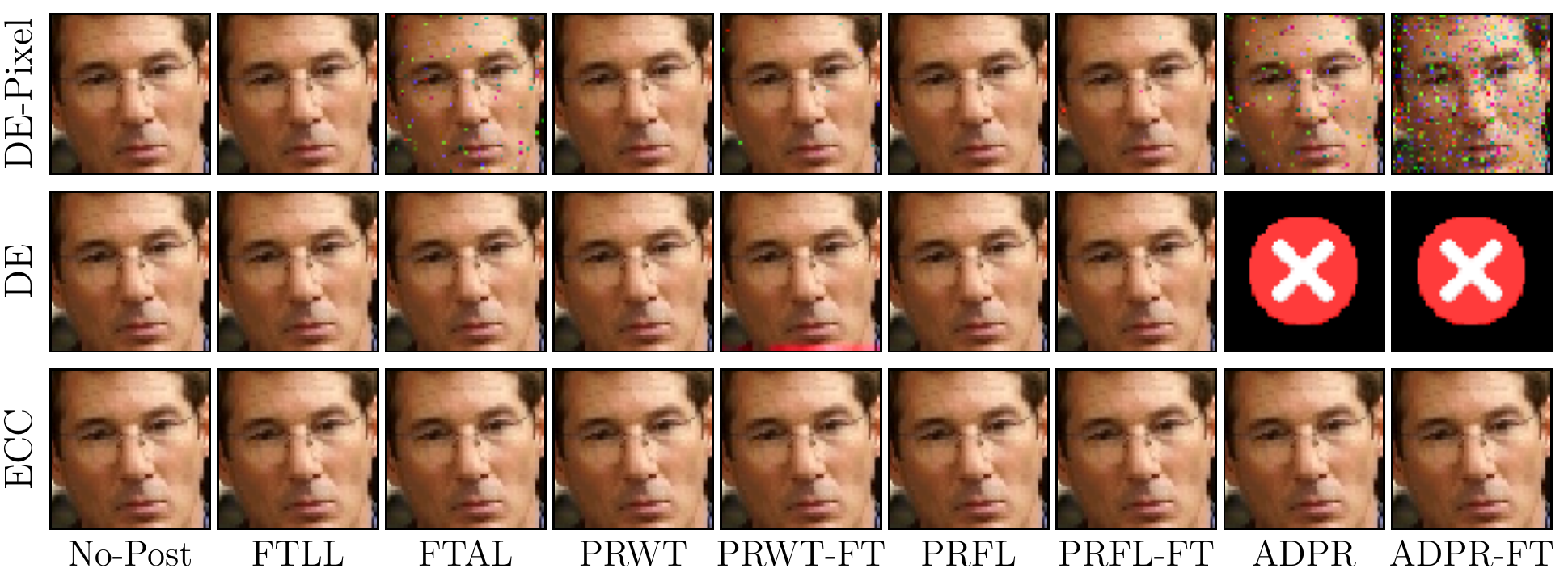}}
    \caption{The recovered face images in training data inference. The 9 columns correspond to no post-processing and the 8 post-processing methods, respectively. The two black images with ``X'' mean that the JPEG files are corrupted.}
    \label{Trainingdata}
    \vspace{-2mm}
\end{figure*}

\subsection{Training Data Inference}
\subsubsection{Experimental Setup} An attacker can treat training data points as a message and embed it into the neural network during training. After the neural network is deployed, the attacker can recover the training data points via querying the neural network. 

\myparatight{Dataset} Following previous work~\cite{song2017machine}, we use the FaceScrub dataset~\cite{ng2014data}, which contains URLs to JPEG images for 530 celebrities (around 200 images per person). When we collected the data, some URLs expired and we were able to download 65,372 images. We perform face recognition on this dataset. In particular, we sample 90\% of images as  training dataset and treat the remaining images as  testing dataset. Following previous work~\cite{song2017machine}, we rescale each image to $50 \times 50$ RGB pixels. We adopt an example neural network in Keras, whose architecture is shown in Table~\ref{architecture}.  We learn model parameters using SGD. The number of epochs is 400 and batch size is 64 (30 data points are sampled from the attack dataset and added to each batch). The initial learning rate is 0.1 and multiplied by $0.1$ in the $250$th and $350$th epochs. The parameter settings are different from those for  CIFAR10 in Section~\ref{exp}, because the FaceScrub and CIFAR10 datasets have different characteristics.

\myparatight{Message} Suppose an attacker aims to infer one training image, which we treat as the message. In particular, we explore two ways to represent an image as a message. One way is to represent each RGB value (ranging from 0 to 255) of each pixel as 8 bits and concatenate the pixel values as the bitstream message, which was adopted by previous work~\cite{song2017machine}. The other way is to  treat the binary JPEG image file as the bitstream message. The messages have 60,000  and 20,672 bits in the two ways, respectively.  The second way leads to a shorter message, but a small amount of bit errors in the recovered bitstream may corrupt the JPEG file.

\subsubsection{Results}
Table~\ref{training_data_table} shows the results of training data inference, and Figure~\ref{Trainingdata} shows the recovered face image in different scenarios. DE-Pixel represents the pixel values as the message, while DE and ECC treat the binary JPEG file as the message. When fine tuning is used in post-processing, we set epochs to $40$ and learning rate to $0.01$ (these parameter settings are different from those for the CIFAR10 dataset in Section~\ref{exp} due to the difference between the FaceScrub and CIFAR10 datasets). ECC outperforms ECC-Pixel, so we omit the results of ECC-Pixel.  There are $530$ classes in the FaceScrub dataset. Therefore, each data point in the attack dataset can encode $9$ bits theoretically. However, we found that encoding $5$ bits per data point is more robust, which we adopted.  We note that Song et al.~\cite{song2017machine} had similar observations. 

Compared to DE-Pixel, DE has smaller MSEs, BERs, and accuracy losses (except FTAL) with smaller number of queries. The reason is that the JPEG file based message is shorter than the pixel based one, and thus DE requires a smaller attack dataset. However, DE fails to recover the face image for two post-processing methods (see Figure~\ref{Trainingdata}), because the bit errors are too large such that the JPEG files are corrupted.  Compared to DE-pixel and DE, ECC can accurately and verifiably recover the face image with comparable accuracy losses and 1-5 times more queries. In particular, ECC has slightly higher accuracies than DE-pixel and DE in some cases, while DE has slightly higher accuracies than ECC in other cases. However, the differences are within $1\%$.

\subsection{Property Inference}

\subsubsection{Experimental Setup} Property inference aims to infer certain properties of the training data of a classifier. In particular, we can represent the properties that an attacker is interested in as the message in an information embedding attack.

 \begin{table*} [!t]\renewcommand{\arraystretch}{1.2}
    \centering
    \caption{Results of property inference. API means Asian-Pac-Islander and AIE means Amer-Indian-Eskimo. The letter 'x' means that the corresponding recovered byte is not an ASCII character. Our ECC can correctly recover all the properties in all considered scenarios, while DE can only correctly recover all the properties in 5 of the considered 9 scenarios.}
    \vspace{-2mm}
    \addtolength{\tabcolsep}{-2pt}
    \begin{tabular}{|c|c|c|c|c|c|c|c|c|c|c|c|c|} \hline
      \multirow{2}{*}{Post-Proc.} &
      \multirow{2}{*}{Attack} &
      \multirow{2}{*}{Ave. age} &
      \multirow{2}{*}{Female (\%)} &
        \multicolumn{5}{|c|}{Race (\%)} &
        \multirow{2}{*}{ACC (\%)} &
       \multirow{2}{*}{\#Queries} &
       \multirow{2}{*}{All correct} \\  \cline{5-9}
      &&&  & White & API & AIE & Black & Other&&& \\
      \hline\hline
       \multirow{2}{*}{No-Post}&DE&39&32.43&85.98&02.97&00.95&09.34&00.77&85.21&256&Y \\ \cline{2-12}
                                         &ECC&39 &32.43 & 85.98& 02.97 &00.95 & 09.34  & 00.77&85.19& 288&Y \\ \hline \hline
        \multirow{2}{*}{FTLL}&DE&39&32.43&85.98&02.97&00.95&09.34&00.77&85.21&256&Y \\ \cline{2-12}
                      &ECC &39 &32.43 & 85.98& 02.97 &00.95 & 09.34  & 00.77& 85.18&288&Y \\ \hline \hline
        \multirow{2}{*}{FTAL}&DE&39&32.43&85.98&02.97&00.95&09.34&00.77&85.12&256&Y \\ \cline{2-12}
                      &ECC &39 &32.43 &  85.98& 02.97 &00.95 & 09.34  & 00.77& 85.19&576&Y \\ \hline \hline
        \multirow{2}{*}{PRWT}&DE&79&x6x6x&x7xx8&xxxx7&xxx9x&09x34&40.77&82.28&256&N \\ \cline{2-12}
                      &ECC &39 &32.43 & 85.98& 02.97 &00.95& 09.34 & 00.77& 82.20&1,458&Y \\ \hline \hline
        \multirow{2}{*}{PRWT-FT}&DE&x9&xxx4x&x7xx8&xxxx7&xxxxx&69x74&4x.x7&85.16&256&N \\ \cline{2-12}
                      &ECC &39 &32.43&  85.98& 02.97&00.95& 09.34& 00.77&85.29&1,458&Y \\ \hline \hline
        \multirow{2}{*}{PRFL}&DE&39&32.43&85.x8&22.97&x0.95&09x34&40.77&82.48&256&N \\ \cline{2-12}
                      &ECC &39 &32.43 &  85.98& 02.97 &00.95 & 09.34  & 00.77& 82.54&576&Y \\ \hline \hline
        \multirow{2}{*}{PRFL-FT}&DE&x9&xx.4x&x5xx8&03.x7&xx.9x&x9xx4&40.xx&85.07&256&N \\ \cline{2-12}
                      &ECC &39 &32.43 &  85.98& 02.97 &00.95 & 09.34  & 00.77& 85.23&873&Y \\ \hline \hline
        \multirow{2}{*}{ADPR}&DE&39&32.43&85.98&02.97&00.95&09.34&00.77&83.35&256&Y \\ \cline{2-12}
                      &ECC &39 &32.43 & 85.98& 02.97 &00.95 & 09.34  & 00.77&82.27&576&Y \\ \hline \hline
        \multirow{2}{*}{ADPR-FT}&DE&39&32.43&85.98&02.97&00.95&09.34&00.77&85.13&256&Y \\ \cline{2-12}
                      &ECC &39 &32.43&  85.98& 02.97&00.95 & 09.34  & 00.77&85.24&576&Y \\ \hline 
    \end{tabular}
    \label{compare_with_cryptographic}
    \vspace{-3mm}
\end{table*}

\myparatight{Dataset} We use the US Census Income dataset~\cite{asuncion2007uci}, which contains 45,222 instances (30,162 training instances and 15,060 testing instances, respectively) with unknown values removed. The dataset contains demographic information such as age, gender, and race. The classification task is to predict whether an individual has a salary greater than \$50k. For simplicity, we use a fully connected neural network with two hidden layers (50 and 50 neurons, respectively).\footnote{Exploring the best neural network architecture is not our focus.} We learn model parameters using SGD, where we set 60 epochs and batch size to 64 (10 data points are sampled from the attack dataset and added into each batch). The initial learning rate is 0.1 and reduced by a factor $0.1$ in the 20th and 50th epochs.

\myparatight{Message} Suppose an attacker is interested in the following properties of the training dataset: average age, fraction of female (may be used to measure gender bias), and race distribution. In particular, the dataset has five categories of races, i.e.,  White, Asian-Pac-Islander (API), Amer-Indian-Eskimo (AIE), Black, and Others. We represent the average age using a two-digit integer (e.g., 39), the fraction of female using a four-digit number with two decimals (e.g., 32.43\%), and each race category using a four-digit number with two decimals. We concatenate the values (average age, fraction of female, and race distribution) as an ASCII character stream and further represent the binary form of the ASCII character stream as the bitstream message. Therefore, the message has $8\times (2 + 5 + 5\times5)$ =256 bits, where the average age has two bytes and fraction of female or each race has 5 bytes. After the bitstream message is recovered, the attacker can segment it and recover the considered properties as the attacker knows the semantics of the bitstream. 

\subsubsection{Results}

Table~\ref{compare_with_cryptographic} shows the results of property inference. When fine-tuning is used, we set 40 epochs and the learning rate to be $0.001$. Our attack ECC can accurately and verifiably recover the properties in all considered scenarios, while the existing attack DE fails to correctly recover some properties for four post-processing scenarios. ECC and DE achieve comparable classification accuracies, i.e., the differences are within $1\%$ except ADPR where ECC has 1.08\% lower accuracy than DE. ECC has higher accuracies than DE in some scenarios even if ECC uses a larger attack dataset, because of the randomness in SGD and stopping post-processing when the accuracy loss is more than 3\%.  
ECC requires 1-5.7 times more queries than DE. The encoding ratio of Turbo code is 5 and  ECC uses a 32-bit checksum, which explain the 5.7 times more queries.

\subsection{Neural Network Architecture Inference}
\subsubsection{Experimental Setup} To infer the neural network architecture, an attacker embeds the architecture information into the neural network classifier during training and recovers it via querying the classifier.  

\myparatight{Dataset} We use the CIFAR10 dataset and  ResNet-18  
as ResNet-18 has a more complex architecture than the other neural networks we evaluated. Moreover, we use the same experimental setting to train the model as we did in Section~\ref{exp}.  

\myparatight{Message} One challenge of using information embedding attacks to infer neural network architectures is how to design a general data structure to represent a neural network architecture. To address the challenge, we leverage the general data structure used by Keras to represent a neural network architecture. 
In particular, Keras provides an API to save a neural network architecture in a text file. One way is to directly treat the text file as the message. However, the text file is very large (33KB) due to redundancy in the data structure, making it more challenging to recover the message in an information embedding attack. We found that the text file has some default strings (e.g., the default strides for convolution2d is (1,1)). Therefore, we first remove the default strings, and then we consider two ways to construct the message. One way is to treat the text file with the default strings removed as the message (10.3KB), and the other way is to compress the text file with the default strings removed as the message. In particular, we use gzip to compress the text file in our experiments, which leads to a message with 0.9KB.

\begin{table*} [!t]\renewcommand{\arraystretch}{1.2}
    \centering
    \caption{Results of neural network architecture inference. Our attack ECC can accurately and verifiably recover the neural network architecture in all considered scenarios with comparable accuracy losses with the existing attack DE and 1-5 times more queries than DE.}
    \vspace{-2mm}
    \begin{tabular}{|c|c|c|c|c|c|c|c|c|c|c|c|c|} \hline
      Metric &
      Attack &
     No-Post & FTLL & FTAL&PRWT&PRWT-FT & PRFL & PRFL-FT & ADPR & ADPR-FT \\
      \hline
	
	        \multirow{3}{*}{BER (\%)}&DE-Raw&0&4.8&16.8&13.8&19.7&10.7&18.0&20.0&22.9 \\ \cline{2-11}
                                  &DE&0&0&0.5&0.6&1.2&0.4&1.2&6.3&1.7 \\ \cline{2-11}
                                         &ECC &0&0&0&0&0&0&0&0&0 \\ \hline

       \multirow{3}{*}{Correct?}&DE-Raw&Y&N&N&N&N&N&N&N&N\\ \cline{2-11}
                                         &DE&Y&Y&N&N&N&N&N&N&N\\ \cline{2-11}
                                         &ECC &Y&Y&Y&Y&Y&Y&Y&Y&Y \\ \hline
              \multirow{3}{*}{ACC (\%)}&DE-Raw&88.75&88.47&89.03&85.75&88.40&86.15&88.48&86.10&88.89 \\ \cline{2-11}
                            &DE&90.14&89.89&90.31&87.27&89.50&87.82&89.49&87.97&89.63 \\ \cline{2-11}
                            &ECC &89.52&89.37&90.04&86.52&89.40&87.09&89.30&87.35&89.46  \\ \hline

       \multirow{3}{*}{\#Queries}&DE-Raw&28,320&28,320&28,320&28,320&28,320&28,320&28,320&28,320&28,320 \\ \cline{2-11}
                                  &DE&2,411&2,411&2,411&2,411&2,411&2,411&2,411&2,411&2,411 \\ \cline{2-11}
                                         &ECC &2,422&2,422&7,268&12,113&12,113&4,843&12,113&7,268&7,268 \\ \hline
    \end{tabular}
    \label{architecture-infer}
    \vspace{-3mm}
\end{table*}

\subsubsection{Results}

Table~\ref{architecture-infer} shows the results of neural network architecture inference. DE-Raw means that we treat the raw text file with the default strings removed as the message, while DE and ECC treat the compressed  file  as the message. For DE-Raw, we sample 20 data points from the attack dataset for each batch during training (10 data points lead to much higher BERs), while we sample 10 data points for DE and ECC as we did in Section~\ref{exp}. Compared to DE-Raw, DE achieves lower BERs, smaller accuracy losses, and less queries. This is because DE uses a shorter message. However, both DE-Raw and DE are not robust against post-processing (DE is only robust against FFLL), i.e., they cannot correctly recover the architecture when the neural network is post-processed. Compared to DE, our attack ECC can correctly recover the neural network architecture in all considered scenarios with comparable accuracy losses (differences are within $1\%$) and 1-5 times more queries.

%% file: related.tex
\section{Related Work}
\label{relatedwork}

\myparatight{Information embedding attacks} Song et al.~\cite{song2017machine} proposed  information embedding attacks, which can embed the messages into model parameters (white-box attack) or labels of an attack dataset (black-box attack). We consider black-box attacks as they are more general. In the black-box attacks proposed by Song et al.,  the Embed function directly encodes the message using the labels of the attack dataset and the Recover function queries all data points in the attack dataset. These attacks are not robust against certain post-processing of the learnt model and are not verifiable. 
Information embedding attacks essentially exploit the large capacity of a neural network, which can remember more information other than the main task it is trained for. Such memorization capacity was demonstrated earlier by Zhang et al.~\cite{zhang2016understanding}.

\myparatight{Data poisoning attacks} Information embedding attacks rely on data poisoning attacks in the Embed function. Roughly speaking, data poisoning attacks aim to manipulate the training phase of machine learning via poisoning the training dataset. Depending on the goals of data poisoning attacks, we can roughly classify data poisoning attacks to three categories, i.e., \emph{untargeted data poisoning attacks},    \emph{targeted data poisoning attacks}, and \emph{backdoor attacks}. Untargeted data poisoning attacks~\cite{rubinstein2009antidote,biggio2012poisoning,biggio2013poisoning,xiao2015feature,Jagielski18,poisoningattackRecSys16,YangRecSys17,fang2018poisoning,fang2020influence,fang2020local,jia2020intrinsic} aim to poison the training dataset such that the learnt classifier has high error rates indiscriminately for testing inputs.   Targeted data poisoning attacks~\cite{Nelson08poisoningattackSpamfilter,KermaniPoisoninghealthcare,Suciu18,shafahi2018poison,ji2018model} aim to poison the training dataset such that the learnt classifier has attacker-desired predictions for a particular set of inputs without affecting the predicted labels for the remaining inputs. Backdoor attacks~\cite{Liu18,Gu17,Chen17,zhang2020backdoor,wang2020certifying} aim to learn a classifier that has attacker-desired predictions for any input with a particular trigger. Information embedding attacks rely on targeted data poisoning attacks, where the learnt classifier predicts attacker-desired labels for the data points in the attack dataset. Liu et al.~\cite{liu2018fine} proposed fine-pruning (denoted as ADPR and ADPR-FT in our experiments) as a defense against backdoor attacks. Fine-pruning could also be used to post-process a classifier to remove the embedded message. Our experimental results show that our information embedding attacks are robust against fine-pruning. 


\myparatight{Membership, property, parameter, and hyperparameter inference attacks} These are basic attacks to compromise confidentiality/privacy in machine learning. For instance, membership inference attacks~\cite{membershipInfer,membershipLocation,Melis19,jia2019memguard} can infer whether a data point is in the training dataset that is used to train a model, which are stronger  than model inversion attacks~\cite{fredrikson2014privacy,fredrikson2015model} and property inference attacks~\cite{Ateniese15,Ganju18,Melis19}. Parameter and hyperparameter inference attacks~\cite{tramer2016stealing,PracticalBlackBox17,WangHyper18} aim to steal the parameters/hyperparameters of a deployed model. These attacks assume the training phase maintains integrity when learning the classifier, and they aim to compromise the confidentiality/privacy via analyzing the learnt classifier under either white-box or black-box setting. Information embedding attacks assume the training phase can be manipulated by the attacker. Information embedding attacks could be applied to perform these privacy attacks, e.g., the message can be neural network architecture for hyperparameter inference attacks.

%% file: conclusion.tex
\section{Conclusion and Future Work}
In this paper, we propose a new information embedding attack that is verifiable and robust against post-processing of the classifier. In particular, our attack leverages CRC to verify the correctness of the recovered message. Moreover, our attack leverages Turbo codes to encode the message before embedding it to the DNN classifier, which is robust against post-processing. We show that we can leverage the property of Turbo codes, which  supports error correcting with a partial code, to adaptively recover the message, which queries the classifier only if needed. We evaluate our attacks using simulated messages and apply our attacks to three applications including training data inference, property inference, and DNN architecture inference. Compared to state-of-the-art attacks, our attacks can verifiably and accurately recover the messages for 8 popular post-processing methods with at most $1\%$ more accuracy loss of the classifier (except one case) and 1-5 times more queries, which are acceptable. 
Interesting directions for future work include 1) designing new defenses against our coding-based information embedding attacks and 2) applying our information embedding attacks to more applications.